\documentclass[twocolumn%
              ,superscriptaddress%
              ,aps%
              ,pra]{revtex4}

\usepackage{amsmath,amssymb}
\usepackage{braket}
\usepackage{graphicx}
\usepackage{hyperref}
\usepackage{bm} % bold math symbols

\begin{document}

\title{Proximity force approximation and specular reflection:
       Application of the WKB limit of Mie scattering to the Casimir effect}

\author{Benjamin Spreng}
\affiliation{Universit{\"a}t Augsburg, Institut f{\"u}r Physik, 86135 Augsburg, Germany}
\author{Michael Hartmann}
\affiliation{Universit{\"a}t Augsburg, Institut f{\"u}r Physik, 86135 Augsburg, Germany}
\author{Vinicius Henning}
\affiliation{Instituto de F{\'i}sica, Universidade Federal do Rio de Janeiro, CP 68528, Rio de Janeiro RJ 21941-909, Brazil}
\author{Paulo A. Maia Neto}
\affiliation{Instituto de F{\'i}sica, Universidade Federal do Rio de Janeiro, CP 68528, Rio de Janeiro RJ 21941-909, Brazil}
\author{Gert-Ludwig Ingold}
\affiliation{Universit{\"a}t Augsburg, Institut f{\"u}r Physik, 86135 Augsburg, Germany}

\date{\today}

\begin{abstract}
The electromagnetic Casimir interaction between two spheres is studied within
the scattering approach using the plane-wave basis. It is demonstrated that
the proximity force approximation (PFA) corresponds to the specular-reflection
limit of Mie scattering. Using the leading-order semiclassical WKB
approximation for the direct reflection term in the Debye expansion for the
scattering amplitudes, we prove that PFA provides the correct leading-order
divergence for arbitrary materials and temperatures in the sphere-sphere
and the plane-sphere geometry. Our derivation implies that only a small
section around the points of closest approach between the interacting spherical
surfaces contributes in the PFA regime. The corresponding characteristic length
scale is estimated from the width of the Gaussian integrand obtained within the
saddle-point approximation. At low temperatures, the area relevant for the
thermal corrections is much larger than the area contributing to the zero-temperature result.
\end{abstract}

\maketitle

\section{Introduction}

The Casimir force between material surfaces \cite{Casimir1948} is a remarkable
prediction of quantum electrodynamics~\cite{Bordag2009}. Precise measurements of
the Casimir force between metallic surfaces are now capable of distinguishing
different models for the metallic
conductivity~\cite{Decca2007,Decca2007EPJC,Sushkov2011,Chang2012}, including the
case of magnetic materials~\cite{Banishev2013,Bimonte2016}. In order to minimize
systematic errors, all such recent experiments, as well as the majority of older
ones \cite{Decca2011,Lamoreaux2011}, probe the Casimir attraction between a
spherical and a planar surface, instead of employing the geometry with plane
parallel plates. In addition, in view of recent experiments
\cite{Ether2015,Garrett2018} the sphere-sphere geometry has gained interest.

On the theoretical side, the scattering
approach~\cite{Lambrecht2006,Emig2007,Rahi2009} allows to compute the Casimir
energy from the scattering matrices of the individual bodies interacting across
a region of empty space. It also provides a clear physical picture of the
Casimir effect as resulting from the reverberation or multiple scattering of
vacuum or thermal electromagnetic field fluctuations between the interacting
surfaces \cite{Jaekel1991,Genet2003}.

These theoretical and experimental advances were disconnected until very
recently, when exact numerical results for typical experimental conditions were
derived from the scattering approach~\cite{Hartmann2017}. Instead of applying
the recent theoretical developments, the surface curvature in real experiments
is taken into account with the help of the proximity force (Derjaguin)
approximation (PFA)~\cite{Derjaguin1934}, in which the result for the Casimir
energy between parallel plates is averaged over the local distances
corresponding to the geometry of interest. PFA is also often employed in
surface science \cite{Butt2010}, for instance in the comparison with
experimental results for the van der Waals interaction between spherical
colloids~\cite{Borkovec2012,Elzbieciak-Wodka2014}.

The PFA approach to spherical curvature is conceptually different from the
picture of electromagnetic field reverberation between spherical surfaces that
results from the scattering approach. Nevertheless, we establish in this paper
a direct connection between the two approaches. In the limit of a vanishing
distance between the interacting surfaces, the dominant contribution to the
multiple scattering between the surfaces is shown to result from semiclassical
WKB specular reflection by a small section of the sphere's surface around the
point of closest approach. The final expression for the Casimir force then
coincides with the PFA result for the leading-order divergence.

The connection between the scattering approach and PFA was first analyzed in a
different context. The roughness correction to the Casimir energy for parallel
planes was derived as a small perturbation of the ideal parallel-plates
geometry~\cite{MaiaNeto2005}. The PFA result for the roughness correction was
then derived from the more general case in the limit of short distances and
smooth surfaces by considering the value of the perturbative kernel at zero
momentum~\cite{MaiaNeto2005}. This line of reasoning can be generalized to the
entire perturbative series~\cite{Fosco2014} within the derivative expansion
approach~\cite{Fosco2011}, although it is not possible to derive explicit
results for all the corresponding kernels in this case. Nevertheless, the PFA
result is obtained in leading order for surfaces that can be continuously
deformed from the planar symmetry, as long as the kernel functions have a well
defined limit at zero momentum~\cite{Fosco2014}. Since the first condition does
not hold for compact objects, this derivation does not constitute a proof of PFA
for the plane-sphere nor for the sphere-sphere geometries, even though the
leading-order correction to PFA has been successfully
derived~\cite{BimonteEPL2012,BimonteAPL2012} whenever the perturbative kernel
is analytical at zero momentum~\cite{Mazzitelli2015}.

For the plane-sphere setup, the validity of PFA in the short-distance limit was
shown at zero temperature for both perfect \cite{Bordag2010,Teo2011}
and real metals \cite{Teo2013} by developing the scattering approach in the
multipolar basis and taking suitable asymptotic approximations for the relevant
spherical functions. Earlier results were derived for a scalar field model at
zero temperature~\cite{Bulgac2006,Bordag2008}. In the opposite limit of high temperatures,
a similar derivation was recently undertaken for perfect metals
\cite{Bimonte2017}. Ref.~\cite{Bimonte2012} derived an exact formula, also
compatible with the PFA leading-order result, for Drude metals in the
high-temperature limit by using bispherical multipoles.

\begin{figure}
 \includegraphics[width=0.4\columnwidth]{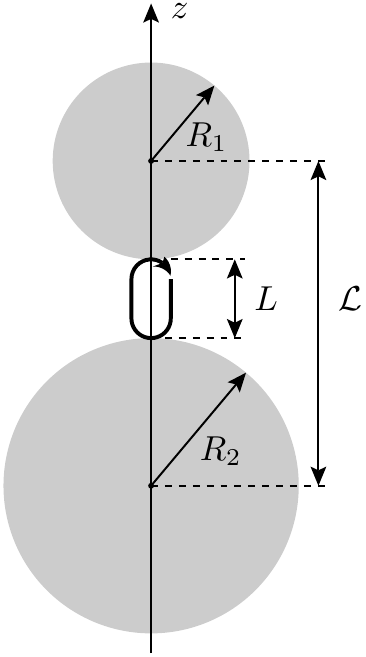}
 \caption{Scattering geometry consisting of two spheres with radii $R_1$ and
          $R_2$ and a surface-to-surface distance $L$. The distance between
          the spheres' centers is $\mathcal{L}=R_1+R_2+L$. The round trip
	  discussed in the text is displayed between the two spheres.}
 \label{fig:geometry}
\end{figure}

Here, we consider the general case of arbitrary temperatures and materials. Our
setup consists of two spheres with radii $R_1$ and $R_2$ in empty
space, at a distance of closest approach $L,$ as indicated in
Fig.~\ref{fig:geometry}. The center-to-center distance is ${\cal L}=L+R_1+R_2$
along the $z$-axis. The plane-sphere case is obtained at any step of our
derivation by taking the radius of one sphere to infinity.

The sphere-sphere Casimir interaction has been analyzed within the scattering
approach for large and moderate distances
\cite{Emig2007,Rodriguez-Lopez2011,Umrath2016,Ether2015}. In addition, for
idealized boundary conditions, the leading-order correction to the PFA result
has been obtained \cite{Teo2012}. In all cases, an expansion in terms of
angular momentum multipolar waves has been employed, which is particularly
natural for not too small distances. In this paper, we show that the linear
momentum representation is better adapted to the short-distance PFA regime $L
\ll R_1, R_2$ because of its direct connection to the physical picture of
specular reflection between the spherical surfaces.

Our derivation brings into light the main physical ingredients underlying the
PFA regime. We show that only the direct reflection term in the Debye expansion
\cite{Nussenzveig69} of the Mie scattering matrix contributes. Moreover, this
contribution is taken in the semiclassical WKB approximation, which has a direct
physical interpretation in terms of specular reflection by the sphere's
surface~\cite{Nussenzveig92}. More importantly, the multiple scattering between
the surfaces defines a scale for the variation of the momentum component
parallel to the $x$-$y$ plane. This scale will allow us to specify the spherical cap
on the spheres that actually contributes in the PFA regime.

Our semiclassical WKB derivation in the momentum representation should not be
confused with the semiclassical treatments of the Casimir interaction in the
position representation \cite{Schaden1998,Jaffe2004,Scardicchio2005}. The
standard Gutzwiller trace formula fails badly in the PFA regime because the
relevant surface area increases as one approaches this limit \cite{Jaffe2004}.
In other words, the semiclassical PFA limit cannot be connected to periodic
orbits obtained from a stationary-phase approximation in the position
representation, because position is poorly resolved in this limit. On the other
hand, the condition of specular reflection makes the integration range for the
conjugate momentum variable increasingly narrow as the distance between the
spheres becomes very small compared to their radii, allowing us to obtain the
PFA result from a saddle-point approximation for the scattering formula.

The paper is organized as follows. Sec.~\ref{sec:planewavebasis} presents the
general development of the scattering formula for the Casimir free energy in the
plane-wave basis.  The following sections apply this formalism to the case of
two spheres in the short-distance limit. The Mie scattering matrix elements in
the plane-wave basis, for arbitrary directions of incidence, are presented in
Sec.~\ref{sec:scattering_at_sphere}. The corresponding WKB approximation is
derived in Sec.~\ref{sec:wkb} and Sec.~\ref{sec:pfa_from_spa} calculates the
resulting Casimir free energy in the saddle-point approximation.
Sec.~\ref{sec:effective_area} presents an estimation of the area on the spheres
that actually contributes in the short-distance limit. Concluding remarks are
presented in Sec.~\ref{sec:conclusions} and the appendices contain some
additional technical details.

\section{The Casimir free energy in the plane-wave basis}
\label{sec:planewavebasis}

In a homogeneous medium, the electric and the magnetic field satisfy the vector
Helmholtz equation. A convenient basis set consists of plane waves
characterized by a wave vector $\mathbf{K}$ and the polarization $p$. In order
to define the polarization basis, we assume a given $z$-axis which can be
appropriately fixed later on and choose an incidence plane spanned by the
$z$-axis and the wave vector $\mathbf{K}$. Denoting unit vectors by a hat, we
obtain the basis vectors for transverse electric (TE) and transverse magnetic
(TM) modes as
\begin{equation}
\label{eq:polarizationTETM}
\hat{\bm{\epsilon}}_\mathrm{TE} = \frac{\hat{\mathbf{z}} \times \hat{\bm{K}}}
                                 {\vert\hat{\mathbf{z}} \times \hat{\bm{K}}\vert}, \quad
\hat{\bm{\epsilon}}_\mathrm{TM} = \hat{\bm{\epsilon}}_\mathrm{TE} \times \hat{\mathbf{K}} \,.
\end{equation}
Thus, the TE polarization is perpendicular to the incidence plane while the TM
polarization lies in it.

As we will see later in this section, it is convenient to fix the frequency which
by means of the dispersion relation is obtained as
$\omega=c\vert\mathbf{K}\vert$ with $c$ being the speed of light. It is then
sufficient to specify the projection $\mathbf{k}=(K_x, K_y, 0)$ of the wave vector
$\mathbf{K}$ onto the $x$-$y$ plane perpendicular to the $z$-axis. In order
to uniquely define the wave vector, we finally need to fix the direction
of the propagation in the $z$-direction by
\begin{equation}
	K_z = \phi k_z, \quad k_z \equiv (\omega^2/c^2-\mathbf{k}^2)^{1/2} \,,
\end{equation}
with $\phi=\pm1$. We thus arrive at the angular spectral representation expressing
the plane-wave basis as $\{\ket{\omega,\mathbf{k},p,\phi}\}$ \cite{Nieto-Vesperinas2006}.
In position space, the basis functions read
\begin{equation}
\braket{x,y,z | \omega,\mathbf{k},p,\phi} = \hat{\bm{\epsilon}}_p
	\left(\frac{1}{2\pi}\left|\frac{\omega}{c k_z}\right|\right)^{1/2}
	e^{i(\mathbf{k}\mathbf{r}+\phi k_z z)}
\end{equation}
where $\mathbf{r}=(x,y,0)$. The normalization factor is appropriate for the
angular spectral representation where the integration is performed over
the frequency and the projection of the wave vector into the $x$-$y$ plane.

For the purpose of this section, we consider two arbitrary objects that can be
separated by a plane parallel to the $x$-$y$ plane. The reference point of
object~1 is located in the origin and the reference point of object~2 is
located at $z=\mathcal{L}$.  The starting point of our analysis is the
scattering approach to the Casimir effect in imaginary frequencies
$\xi=i\omega$ where the free energy is expressed as
\cite{Lambrecht2006,Emig2007}
\begin{equation}
\label{eq:F}
\mathcal{F} = \frac{k_\mathrm{B} T}{2} \sum_{n=-\infty}^\infty
              \mathrm{tr}\log\left[1-\mathcal{M}(\vert\xi_n\vert)\right]
\end{equation}
in terms of a sum over the Matsubara frequencies $\xi_n=2\pi n k_\mathrm{B}
T/\hbar$. The central object here is the round-trip operator
\begin{equation}
\label{eq:round-trip}
\mathcal{M} = \mathcal{R}_1\mathcal{T}_{12}\mathcal{R}_2\mathcal{T}_{21}
\end{equation}
describing a complete round trip of an electromagnetic wave between the two
scatterers in the order indicated in Fig.~\ref{fig:geometry}. $\mathcal{R}_j$
denotes the reflection operator at object $j=1,2$ and $\mathcal{T}_{21}$
describes a translation from the reference frame of object 1 to the reference
frame of object 2, and vice versa for $\mathcal{T}_{12}$.

In the plane-wave basis the translation operators are diagonal with matrix
elements $e^{-\kappa\mathcal{L}}$ where $\kappa = (\xi^2/c^2+k^2)^{1/2}$ denotes
the $z$-component of the wave vector associated with the imaginary frequency
$\xi$. As the frequency $\xi$ remains constant during a round trip, we
suppress the notation $\xi$ in the labeling of the basis elements.

The logarithm in \eqref{eq:F} can be expanded in a Mercator series
\begin{equation}
\label{eq:F_roundtrip}
\mathcal{F} = -\frac{k_\mathrm{B}T}{2} \sum_{n=-\infty}^\infty \sum_{r=1}^\infty \frac{1}{r} \mathrm{tr}\mathcal{M}^r\left(\left|\xi_n\right|\right)
\end{equation}
and the trace of the $r$-th power of the round-trip operator is given by
\begin{equation}
\begin{aligned}
\mathrm{tr}\mathcal{M}^r &= \sum_{p_1,\dotsc,p_{2r}} \int \frac{d \mathbf{k}_1\dots d\mathbf{k}_{2r}}{(2\pi)^{4r}}
\prod_{j=1}^r e^{-(\kappa_{2j}+\kappa_{2j-1})\mathcal{L}} \\
\label{eq:M_elems}
&\qquad\times \braket{\mathbf{k}_{2j+1}, p_{2j+1}, - \vert\mathcal{R}_1 \vert\mathbf{k}_{2j}, p_{2j}, +} \\
&\qquad\times\braket{\mathbf{k}_{2j}, p_{2j}, + \vert\mathcal{R}_2 \vert\mathbf{k}_{2j-1}, p_{2j-1}, -} \,.
\end{aligned}
\end{equation}
Here, we have used the convention of cyclic indices $p_{2r+1}\equiv p_1$ and
$\mathbf{k}_{2r+1}\equiv\mathbf{k}_1$. Eqs.~\eqref{eq:F_roundtrip} and
\eqref{eq:M_elems} can be interpreted as an expansion in round trips. The free
energy consists of contributions from a single round trip within the cavity, up
to infinitely large numbers of round trips.  Also, the expansion in round trips
is a natural way to compute $\mathrm{tr}\log\left(1-\mathcal{M}\right)$ of a
non-diagonal round-trip operator $\mathcal{M}$ expressed in a continuous basis.

The exponential factor in \eqref{eq:M_elems} might suggest that the component in
$z$-direction $\kappa$ of the imaginary wave vectors is constrained to values of
the order of $1/\mathcal{L}$. However, we will see later for the special case of
the sphere-sphere configuration that the reflection matrix elements grow
exponentially with the size of the corresponding objects. As a consequence,
much larger values for $\kappa$ of the order of $1/L$ are possible, where $L$
denotes the closest distance between the two objects.

While our discussion so far was fairly general, we will specialize on the
geometry of two spheres in the following sections.

\section{Scattering at a sphere}
\label{sec:scattering_at_sphere}

While the translation operators in the plane-wave basis are trivial, the
reflection operators require more care. For scattering at a single sphere, we
can make the scattering plane spanned by the initial and the reflected wave
vectors coincide with the incidence plane defining the polarizations as
explained in Sect.~\ref{sec:planewavebasis}. In the sphere-sphere setup shown
in Fig.~\ref{fig:geometry}, however, the $z$-axis is defined by the centers of
the two spheres. Then, in general the scattering plane and the incidence plane
will not coincide and one has to change the polarization basis as explained in
appendix~\ref{sec:appendixA}. Nevertheless, it will turn out that the results
can mostly be cast in quantities familiar from the standard Mie theory,
allowing us to relate PFA to the concepts of geometrical optics.

\begin{figure}
\includegraphics[width=0.7\columnwidth]{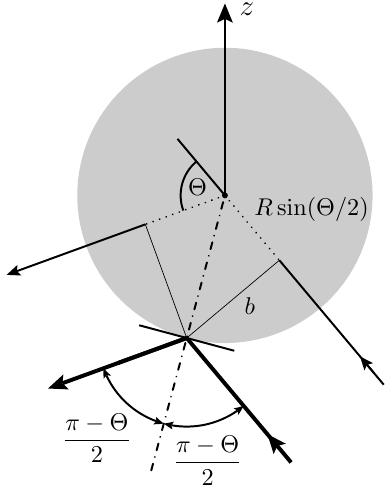}
\caption{Geometrical optics limit for the direct reflection by a sphere of
         radius $R$. Within the WKB approximation, a given scattering angle
         $\Theta$ defines the impact parameter $b=R\cos(\Theta/2).$ Seen from
         the tangent plane to the sphere, the angle of incidence is given
         by $(\pi-\Theta)/2$. The missing phase of a ray with frequency $\omega$
         reflected on the sphere's surface with respect to a corresponding ray
         passing via the sphere's center before being deflected amounts to
	     $2(\omega R/c)\sin(\Theta/2)$.}
\label{fig:phaseshift}
\end{figure}

Then, the matrix elements of the reflection operator $\mathcal{R}$ at a sphere are
given by
\begin{equation}
\label{eq:RS}
 \begin{aligned}
 \braket{\mathbf k_j, \mathrm{TM} | \mathcal{R} | \mathbf k_i, \mathrm{TM}} &= \phantom{-}\frac{2\pi c}{\xi \kappa_j} \big(A S_2(\Theta)+B S_1(\Theta)\big) \\
 \braket{\mathbf k_j, \mathrm{TE} | \mathcal{R} | \mathbf k_i, \mathrm{TE}} &= \phantom{-}\frac{2\pi c}{\xi \kappa_j} \big(A S_1(\Theta)+B S_2(\Theta)\big) \\
 \braket{\mathbf k_j, \mathrm{TM} | \mathcal{R} | \mathbf k_i, \mathrm{TE}} &=           -\frac{2\pi c}{\xi \kappa_j} \big(C S_1(\Theta)+D S_2(\Theta)\big) \\
 \braket{\mathbf k_j, \mathrm{TE} | \mathcal{R} | \mathbf k_i, \mathrm{TM}} &= \phantom{-}\frac{2\pi c}{\xi \kappa_j} \big(C S_2(\Theta)+D S_1(\Theta)\big)\,,
 \end{aligned}
\end{equation}
where the prefactor results from the normalization within the angular spectral
representation. Here, we have omitted the value of $\phi$ which should be
different for the two waves involved in a matrix element. In addition, the
signs of $C$ and $D$ depend on the direction of propagation as specified in
Eq.~(\ref{eq:abcd}).  The Mie scattering amplitudes for polarizations
perpendicular and parallel to the scattering plane are given by \cite{BH}
\begin{equation}
\label{eq:SA}
\begin{aligned}
S_1(\Theta) &= \sum_{\ell=1}^\infty \frac{2\ell+1}{\ell(\ell+1)}
    \left[ a_\ell\pi_\ell\big(\cos(\Theta)\big) + b_\ell\tau_\ell\big(\cos(\Theta)\big)\right]\\
S_2(\Theta) &= \sum_{\ell=1}^\infty \frac{2\ell+1}{\ell(\ell+1)}
    \left[ a_\ell\tau_\ell\big(\cos(\Theta)\big) + b_\ell\pi_\ell\big(\cos(\Theta)\big)\right]\,,
\end{aligned}
\end{equation}
respectively. The scattering angle is defined relative to the forward direction
(cf.\ Fig.~\ref{fig:phaseshift}) and, for imaginary frequencies, is given by
\begin{equation}
\cos(\Theta) = -\frac{c^2}{\xi^2} \left(\mathbf{k}_i\cdot\mathbf{k}_j + \kappa_i\kappa_j\right) \,.
\label{eq:costheta}
\end{equation}
The angular functions $\pi_\ell$ and $\tau_\ell$ are defined by \cite{BH}
\begin{equation}
\begin{aligned}
\pi_\ell(z) &= {P_\ell}^\prime(z) \\
\tau_\ell(z) &= -(1-z^2){P_\ell}^{\prime\prime}(z)+z{P_\ell}^\prime(z)
\end{aligned}
\label{eq:pi_tau}
\end{equation}
with the Legendre polynomials $P_\ell$ and the prime denoting a derivative with
respect to the argument $z$.

The coefficients $A$, $B$, $C$, and $D$ are functions of $\mathbf{k}_i$ and
$\mathbf{k}_j$. Explicit expressions are derived in appendix~\ref{sec:appendixA}
and given in (\ref{eq:abcd}). The Mie coefficients $a_\ell$ and $b_\ell$
\cite{BH} represent the partial wave electric and magnetic multipole scattering
amplitudes, respectively, for an isotropic sphere. They depend on the
electromagnetic response of the sphere material. For simplicity, we restrict
ourselves to homogeneous non-magnetic spheres in the following.

\section{WKB approximation}
\label{sec:wkb}

In order to obtain PFA as the leading asymptotics for large radii $R_1,R_2\gg
L$,  asymptotic expressions for the matrix elements are required.  For this
purpose, it is convenient to write the scattering amplitudes by means of the
Debye expansion \cite{Nussenzveig92}, i.e., a decomposition into an infinite
series of terms representing multiple internal reflections. In the limit of a
large radius $R,$ the direct reflection term of the Debye expansion gives the
main contribution since the phase factor acquired by propagation inside the
spheres yields exponentially small terms when considering the imaginary frequency domain.

For real frequencies and large size parameters $\omega R/c \gg 1$, the
asymptotic expression for the direct reflection term has been derived from the
WKB approximations for the  Mie coefficients and angular functions by taking the
saddle-point approximation for the integral over angular momenta
\cite{Nussenzveig69}. The resulting expression is valid for all scattering
directions except near the forward one, which is not relevant for the Casimir
interaction.  In fact, the Casimir energy as given by
(\ref{eq:F_roundtrip}) and (\ref{eq:M_elems}) is obtained from round trips
containing only backward scattering channels.  Instead of real frequencies, we
need the asymptotics for imaginary frequencies $\xi,$ for which a very similar
WKB derivation can be performed. The resulting expression coincides with  the
one obtained for real frequencies $\omega$ after replacing $\omega\to i\xi.$ The
leading asymptotics of the scattering amplitude is then given by
\begin{equation}\label{eq:SA-asymptotics}
S_p(\Theta) \simeq \frac{1}{2}(\xi R/c) r_p\bigl((\pi-\Theta)/2\bigr)e^{2(\xi R/c) \sin(\Theta/2)}\,,
\end{equation}
with $p=1,2$ corresponding to TE and TM modes, respectively. $r_\mathrm{TE}$
and $r_\mathrm{TM}$ are the familiar Fresnel reflection coefficients
\cite{Landau84} for a wave in vacuum impinging at an angle of incidence
$\theta$ on a medium with permittivity $\varepsilon$
\begin{equation}\label{eq:Fresnel-coeff}
 \begin{aligned}
 r_\mathrm{TE}(\theta) &= \frac{\cos(\theta) - \sqrt{\varepsilon-\sin^2(\theta)}}%
                              {\cos(\theta) + \sqrt{\varepsilon-\sin^2(\theta)}}\,,\\
 r_\mathrm{TM}(\theta) &= \frac{\varepsilon\cos(\theta) - \sqrt{\varepsilon-\sin^2(\theta)}}%
                              {\varepsilon\cos(\theta) + \sqrt{\varepsilon-\sin^2(\theta)}}\,.
 \end{aligned}
\end{equation}

The asymptotics of the scattering amplitudes can be understood in terms of
geometrical optics in the real frequency domain \cite{Nussenzveig65}. For a
given scattering angle $\Theta,$ the main contribution to $S_1$ and $S_2$ in
Eq.~(\ref{eq:SA}) comes from the neighborhood of the angular momentum value
$\ell = (\omega R/c) \cos(\Theta/2)$ \cite{Nussenzveig65}.  In the
semiclassical approximation, the localization principle \cite{Nussenzveig92}
connects waves with  angular momentum $\ell\gg 1$ to localized rays defining an
impact parameter $b=(c/\omega)\ell$. Thus, the derivation of the WKB
approximation (\ref{eq:SA-asymptotics}) defines rays corresponding to the impact
parameter $b=R\cos(\Theta/2)$ shown in Fig.~\ref{fig:phaseshift}.  Such rays hit
the sphere surface with an incidence angle of $(\pi-\Theta)/2,$ which is
precisely the value required for obtaining the scattering angle $\Theta$ from
the condition of specular reflection at the tangent plane indicated in the
figure.  Comparing the reflection at the tangent plane (thick lines) and at the
sphere with its center as reference point, one finds a difference in path length
amounting to $2(\omega R/c)\sin(\Theta/2)$. In this way, the last two factors of
(\ref{eq:SA-asymptotics}) find their natural explanation. The first factor is
responsible for providing the correct scattering cross section proportional to
$R^2$.

The asymptotics of the Mie scattering amplitudes \eqref{eq:SA-asymptotics} does
not cover the zero frequency case, which is required in the Matsubara sum. In
appendix~\ref{sec:appendixB}, we show that the scattering amplitudes for
$\xi=0$ coincide with the scattering amplitudes at finite imaginary frequencies
(\ref{eq:SA-asymptotics}) evaluated at $\xi=0$.

In order to derive the leading asymptotic expression for the scattering matrix elements
(\ref{eq:RS}), we make use of (\ref{eq:SA-asymptotics}) to obtain
\begin{equation}
\label{eq:RS_WKB}
\braket{\mathbf k_j, p_j | \mathcal{R} | \mathbf k_i, p_i} \simeq \frac{\pi R}{\kappa_j} e^{2(\xi R/c)\sin(\Theta/2)} \rho_{p_j,p_i}
\end{equation}
with
\begin{equation}
\label{eq:rhos}
\begin{aligned}
\rho_{\mathrm{TM},\mathrm{TM}} &= \phantom{-}A r_\mathrm{TM} + B r_\mathrm{TE}, \\
\rho_{\mathrm{TE},\mathrm{TE}} &= \phantom{-}A r_\mathrm{TE} + B r_\mathrm{TM}, \\
\rho_{\mathrm{TM},\mathrm{TE}} &=           -C r_\mathrm{TE} - D r_\mathrm{TM}, \\
\rho_{\mathrm{TE},\mathrm{TM}} &= \phantom{-}C r_\mathrm{TM} + D r_\mathrm{TE} \,.
\end{aligned}
\end{equation}
The WKB expression \eqref{eq:RS_WKB} for the reflection matrix element already
indicates an exponential growth with the sphere radius $R$ as anticipated in the
discussion at the end of Sec.~\ref{sec:planewavebasis}.

\section{PFA from saddle-point approximation}
\label{sec:pfa_from_spa}

We will now derive the proximity force approximation based on the scattering
matrix elements (\ref{eq:RS_WKB}) obtained within the WKB approximation. The
main step consists in evaluating within the saddle-point approximation the
trace over $\mathcal{M}^r$ appearing in the expansion (\ref{eq:F_roundtrip}) of
the free energy. This approach requires large sphere radii $R_1, R_2\gg L$
(cf.\ Fig.~\ref{fig:geometry}), a limit in which PFA is expected to hold. While
we will carry out the calculation for the sphere-sphere geometry, we will
briefly comment on the limit of the plane-sphere geometry when appropriate.

\subsection{Round trips within WKB approximation}

The main quantity in the round-trip expansion of the free energy
(\ref{eq:F_roundtrip}) is the trace over the $r$-th power of the round-trip
operator, which in the plane-wave representation is given by (\ref{eq:M_elems}).
After inserting the WKB scattering matrix elements (\ref{eq:RS_WKB}) and
employing polar coordinates $(k_i, \varphi_i)$ in the $x$-$y$ plane, we
express the result in a form suitable for the saddle-point approximation as
\begin{equation}\label{eq:trMr-start}
\begin{aligned}
\mathrm{tr}\mathcal{M}^r &\simeq \int_0^\infty d^{2r} k
\int_0^{2\pi} d^{2r}\varphi\,
g(\mathbf{k}_1,\dots,\mathbf{k}_{2r})\\
&\qquad\qquad\times e^{-(R_1+R_2) f(\mathbf{k}_1,\dots,\mathbf{k}_{2r})}\,.
\end{aligned}
\end{equation}
The symbol $\simeq$ indicates that the result is only valid in the small
distance limit $L \ll R_1, R_2$. Here, we have used $R_1+R_2$ as large
parameter for the saddle-point approximation. Another choice, e.g. an
individual radius, would equally be possible and would yield the same
final result.

In (\ref{eq:trMr-start}) we introduced the function
\begin{widetext}
\begin{equation}
g(\mathbf{k}_1,\dots,\mathbf{k}_{2r}) = \left(\frac{R_1 R_2}{16\pi^2}\right)^r \sum_{p_1,\dotsc,p_{2r}}\prod_{j=1}^r \frac{k_{2j}k_{2j-1}}{\kappa_{2j}\kappa_{2j-1}}\rho^{(1)}_{p_{2j+1},p_{2j}}(\mathbf{k}_{2j+1},\mathbf{k}_{2j}) \rho^{(2)}_{p_{2j},p_{2j-1}}(\mathbf{k}_{2j},\mathbf{k}_{2j-1})e^{-(\kappa_{2j}+\kappa_{2j-1})L}
\label{eq:gfunc}
\end{equation}
\end{widetext}
where the superscript of the factors $\rho$ defined in (\ref{eq:rhos})
indicates the sphere for which the Fresnel reflection coefficient
is to be taken.

The function in the exponent of (\ref{eq:trMr-start}) is given by
\begin{equation}
f(\mathbf{k}_1,\dots,\mathbf{k}_{2r}) = \frac{1}{R_1+R_2}\sum_{j=1}^{r}\left(R_1\eta_{2j} + R_2\eta_{2j-1} \right)
\label{eq:func_f}
\end{equation}
where the terms with even and odd indices are contributions from sphere 1 and 2, respectively, and
\begin{multline}
\label{eq:eta}
\eta_i = \kappa_i + \kappa_{i+1} \\
- \sqrt{2\big[(\xi/c)^2 + \kappa_i\kappa_{i+1} + k_i k_{i+1}\cos(\varphi_{i}-\varphi_{i+1})\big]}\,.
\end{multline}
While the last term on the right-hand side arises from the phase shift
illustrated in Fig.~\ref{fig:phaseshift}, the first two terms are associated
with a translation over twice the radius of the sphere at which the reflection
occurs. As a consequence, the last factor in (\ref{eq:gfunc}) only depends on
the closest distance $L$ between the two spheres.

\subsection{Saddle-point manifold}
\label{subsec:spmanifold}

In order to evaluate the $4r$ integrals in (\ref{eq:trMr-start}) within the
saddle-point approximation, we need to determine the stationary points. In
fact, there exists a two-dimensional manifold of saddle points
\begin{equation}
k_1=\dots=k_{2r}=k_\ast,\quad \varphi_1=\dots=\varphi_{2r}=\varphi_\ast
\label{eq:spmanifold}
\end{equation}
parametrized by $k_\ast$ and $\varphi_\ast$.  Thus, on the saddle-point
manifold, the change of angle $\varphi=\varphi_{j+1}-\varphi_j$ vanishes and
leads to a significant simplification because then the incidence and
scattering planes coincide. As a consequence, $A=1, B=C=D=0$ as can be
verified also from the relations (\ref{eq:abcd}) by setting $\varphi=0$.
Thus, in view of (\ref{eq:rhos}), the polarization is always conserved during
the scattering processes within the saddle-point approximation. The trace
over $r$ round-trip operators (\ref{eq:trMr-start}) can now be decomposed into
two independent polarization contributions
\begin{equation}
\mathrm{tr}\mathcal{M}^r = \mathrm{tr}\mathcal{M}_\mathrm{TE}^r+\mathrm{tr}\mathcal{M}_\mathrm{TM}^r\,.
\end{equation}

The saddle-point manifold (\ref{eq:spmanifold}) also implies that the projection of
the wave vector onto the $x$-$y$ plane is conserved during the reflection. Within
the WKB approximation, this is the case when in Fig.~\ref{fig:phaseshift} the
tangent plane on which the reflection occurs were perpendicular to the
$z$-axis. Under this condition, the WKB phase shift upon reflection $2(\xi
R/c)\sin(\Theta/2)$ can be expressed as $2\kappa R$. This precisely cancels
the term arising from the translation by twice the sphere radius. As a consequence
the exponent (\ref{eq:func_f}) vanishes on the saddle-point manifold,
\begin{equation}
f\big\vert_\text{S.P.} = 0 \,.
\end{equation}

For the prefactor in the integrand of (\ref{eq:trMr-start}), we now obtain
for the two polarization contributions $p=\mathrm{TE},\mathrm{TM}$
\begin{equation}
g_p\big\vert_\text{S.P.}(k_\ast) = \left(\frac{R_1R_2}{16\pi^2}
	\frac{k_\ast^2}{\kappa_\ast^2}r_p^{(1)} r_p^{(2)}\right)^r e^{-2r\kappa_\ast L}
\label{eq:gp}
\end{equation}
on the saddle-point manifold. Here, we have introduced $\kappa_\ast =
(\xi^2/c^2+k_\ast^2)^{1/2}$. The Fresnel coefficients (\ref{eq:Fresnel-coeff})
are evaluated at the angle
\begin{equation}
\theta=\arccos(\kappa_\ast c/\xi)\,.
\label{eq:theta}
\end{equation}
On the saddle-point manifold, only a translation by $L$ remains, which just
corresponds to the distance between the two tangent planes perpendicular to the
$z$-axis facing each other. Thus the relevant length scale for the $z$
component of the imaginary wave vector is given by $L$ instead of $\mathcal{L}$
as \eqref{eq:M_elems} had seemed to imply.

The result (\ref{eq:gp}) might raise some concerns about a divergence
associated with the plane-sphere limit. Choosing without loss of generality
$R_1\leq R_2$, the plane-sphere limit reads $R_2\to\infty$. We know that the
wave vector $\mathbf{k}$ is conserved during a reflection at the plane, so that
$2r$ integrations have to drop out in this limit. In fact, we will see in the
next subsection that for $R_2\to\infty$ indeed $2r$ delta functions appear for
which the factor $R_2/4\pi$ contained in (\ref{eq:gp}) provides the
normalization.

\subsection{Hessian matrix}

We now turn to the Hessian matrix $\textsf{H}$ of the function $f$
(\ref{eq:func_f}) evaluated on the saddle-point manifold. It is found to be
of block-diagonal form
\begin{equation}
\textsf{H} = \begin{pmatrix} \textsf{H}_{kk} & 0\\ 0 & \textsf{H}_{\varphi\varphi} \end{pmatrix}
\end{equation}
with
\begin{equation}
\textsf{H}_{kk} = \frac{1}{2 \kappa_\ast } \textsf{M}_r\,,
\quad  \textsf{H}_{\varphi\varphi} = \frac{k_\ast^2}{2 \kappa_\ast } \textsf{M}_r\,.
\end{equation}
Apart from prefactors, both blocks are given by the $2r\times 2r$ matrix
\begin{equation}
\textsf{M}_r = \begin{pmatrix}
1 & -(1-\mu) &&& -\mu\\
-(1-\mu) & 1 & -\mu && \\
& -\mu & \ddots & \ddots &\\
& & \ddots & \ddots & -(1-\mu)\\
-\mu& & & -(1-\mu) & 1 \\
\end{pmatrix}\,,
\end{equation}
where all empty entries should be set to zero. Here, $\mu=R_1/(R_1+R_2)$ and in
the limit of the plane-sphere geometry, $R_2\to\infty$, $\mu$ goes to zero.
The off-diagonal matrix elements alternate between $-(1-\mu)$ and $-\mu$ in
accordance with the alternating reflection at the two spheres during round
trips. In the special case of a single round trip, $r=1$, the two different
off-diagonal matrix elements add up to yield
\begin{equation}
\textsf{M}_1 = \begin{pmatrix}
1 & -1 \\
-1 & 1
\end{pmatrix}\,.
\end{equation}

The eigenvalues of the matrix $\mathsf{M}_r$ are found as
\begin{equation}\label{eq:eigenvals}
\lambda^{(j)}_\pm = 1 \pm \sqrt{1-4\mu(1-\mu)\sin^2\left(\frac{\pi j}{r}\right)}
\end{equation}
for $j=0,\dots,r-1$.
Both, $\mathsf{H}_{kk}$ and $\mathsf{H}_{\varphi\varphi}$ have a zero eigenvalue
corresponding to the saddle-point manifold discussed in the previous subsection.
Within the saddle-point approximation, the directions perpendicular to the
saddle-point manifold can now be integrated out in the usual way, while the
integration along the families has to be carried out exactly. This can be done
for $\varphi_\ast$ so that we are left with an integral over $k_\ast$ in the
following subsection.

Before turning to the result for the trace over $r$ round-trip operators, we
would like to make another comment on the plane-sphere case. Considering the
eigenvalues (\ref{eq:eigenvals}) to leading order in $\mu$, we find  $r$ eigenvalues
satisfying
\begin{equation}\label{R1R2}
(R_1+R_2)\,\lambda^{(j)}_- = 2R_1\sin^2(\pi j/r)\;\;\;\;\;(\mu\to 0)
\end{equation}
and $r$
eigenvalues $\lambda^{(j)}_+=2.$
 When multiplied by the prefactor $R_1+R_2$ in the exponent of
(\ref{eq:trMr-start}), we obtain from the latter $2r$
delta functions in the limit $R_2\to\infty,$
thus enforcing the conservation of the wave vector $\mathbf{k}$
for the reflection at a plane, as already indicated in the previous subsection.
The
remaining integrals are controlled by (\ref{R1R2}) and lead to the PFA result for the plane-sphere geometry proportional to $R_1.$
Such result can also be obtained from the more general expression for the sphere-sphere case derived in the remainder of this section.

\subsection{Casimir free energy and force}

The evaluation of the saddle-point integral is simplified by
first forming products $\lambda^{(j)}_+\lambda^{(j)}_-$ of the eigenvalues
(\ref{eq:eigenvals}) for $j=1,\ldots,r-1$
and noting that
\begin{equation}
\prod_{j=1}^{r-1} \sin\left(\frac{\pi j}{r}\right) = \frac{r}{2^{r-1}}\,.
\label{eq:sineprod}
\end{equation}
Then, the inverse of the square root of all non-vanishing eigenvalues
of $(R_1+R_2)\textsf{H}$ is found to read
\begin{equation}
\left(\prod_{\lambda\neq0}\lambda\right)^{-1/2} =
\frac{R_\text{eff}}{4r^2}\frac{k_\ast}{\kappa_\ast}
\left(\frac{4\kappa_\ast^2}{k_\ast^2}\frac{1}{R_1R_2}\right)^r\,,
\end{equation}
where we have defined the effective radius
\begin{equation}\label{Reff}
R_\text{eff} = \frac{R_1R_2}{R_1+R_2}\,.
\end{equation}
Changing to the eigenbasis of $\textsf{H}$ but keeping $k_\ast$ and
$\varphi_\ast$ as variables for the integration, i.e., not normalizing the
eigenvectors corresponding to the saddle-point manifolds, yields a factor $2r$
arising through the Jacobian. Then, by applying the multi-dimensional
saddle-point integration formula, we obtain
\begin{equation}\label{Mrp}
\mathrm{tr} \mathcal{M}^r_p \simeq
\frac{R_\text{eff}}{2r}\int_{\vert\xi_n\vert/c}^\infty d\kappa_\ast
\left[r^{(1)}_p r^{(2)}_p e^{-2\kappa_\ast L}\right]^{r}
\end{equation}
for the two polarization contributions $p=\mathrm{TE},\mathrm{TM}$.
Inserting this result into (\ref{eq:F_roundtrip}), we can evaluate the
sum over the number $r$ of round trips and obtain for the free energy
\begin{multline}
\mathcal{F} \simeq -\frac{k_BTR_\text{eff}}{4}\sum_{n=-\infty}^{+\infty}
\sum_{p\in\{\mathrm{TE},\mathrm{TM}\}} \int_{\vert\xi_n\vert/c}^\infty d\kappa_\ast\\
\times\text{Li}_2\left(r_p^{(1)}r_p^{(2)}e^{-2\kappa_\ast L}\right)
\label{eq:F_pfa}
\end{multline}
where $\text{Li}_2$ denotes the dilogarithm \cite{DLMF25.12}. The
Fresnel coefficients (\ref{eq:Fresnel-coeff}) are evaluated at the angle $\theta$ defined in
(\ref{eq:theta}) taken at the frequencies $\vert\xi_n\vert$.

The Casimir force can now be obtained by taking the negative derivative of the
expression (\ref{eq:F_pfa}) for the free energy with respect to the distance
$L$. We thus find the Lifshitz formula
\begin{equation}\label{eq:PFA-result}
F \simeq 2 \pi R_\mathrm{eff} \mathcal{F}_{\rm PP}(L, T)
\end{equation}
with the free energy per area for two planes at distance $L$ and temperature $T$
\begin{multline} \label{FPPLifshitz}
\mathcal{F}_{\rm PP}(L, T) =
\frac{k_\mathrm{B} T}{2}
\sum_{n=-\infty}^{+\infty} \sum_{p\in\{\mathrm{TE},\mathrm{TM}\}} \int_{\vert\xi_n\vert/c}^\infty
\frac{d\kappa}{2\pi} \kappa \\
\times\log\left(1-r^{(1)}_p r^{(2)}_p e^{-2\kappa L} \right)\,.
\end{multline}
It is straightforward to extend this result to the zero temperature case.
As already discussed at the end of the previous subsection, these results
are also valid in the plane-sphere case where $R_\text{eff}$ is replaced
by the sphere radius.

\section{Effective area}
\label{sec:effective_area}

The most precise Casimir experiments employ spherical lenses \cite{Sushkov2011}
or coated microspheres attached to a cantilever beam
\cite{Decca2007,Chang2012,Banishev2013,Garrett2018} instead of whole spherical
surfaces. Since the experimental data are analyzed with the help of the PFA, it
is important to understand what section of the spherical surface actually
contributes to the leading asymptotics. For instance, in the case of a spherical
lens, such analysis would define the minimum transverse lens size required for
equivalence with a complete spherical surface. Here, we estimate the size of the
relevant sphere section and proceed in two steps. First, we employ our
saddle-point calculation to estimate the typical change in the projection of the
wave vector onto the $x$-$y$ plane during reflection at one of the spheres.
Second, we use geometric arguments to obtain the corresponding size of the
sphere section in real space.

\begin{figure}
\includegraphics[width=0.7\columnwidth]{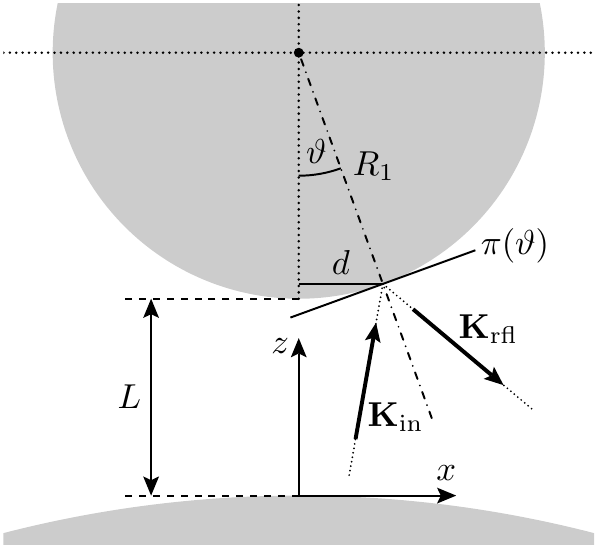}
\caption{Estimation of the effective area contributing to the Casimir interaction
     between two spheres. ${\bf K}_\text{in}$ and ${\bf K}_\text{rfl}$ denote the incident
	 and reflected wave vectors, respectively. To be definite, the reflection is
     shown at the smaller sphere of radius $R_1$. Specular reflection at the tangent plane
     $\pi(\vartheta)$ entails that the projections of ${\bf K}_\text{in}$ and
     ${\bf K}_\text{rfl}$ on $\pi(\vartheta)$ are equal. On the other hand, the wave
     vector projections on the $x$-$y$ plane are generally different:
     $\delta k=|{\bf k}_\text{rfl}-{\bf k}_\text{in}|\approx2\vartheta k_z$ for $\vartheta\ll 1$.
     We can estimate the angular sector effectively contributing to the Casimir interaction
     from the width $\delta k$ of the Gaussian integrand in the saddle-point
	 approximation (see text).
}
\label{fig:effectiveArea}
\end{figure}

Even though we first consider the reflection at a single sphere, this reflection
is still to be taken in the context of the sphere-sphere setup. Therefore, we
keep the saddle-point manifold (\ref{eq:spmanifold}) obtained in
Sec.~\ref{subsec:spmanifold}. Considering only a single reflection, we denote
the incident and reflected wave vectors as $\mathbf{K}_\text{in}$ and
$\mathbf{K}_\text{rfl}$, respectively, as indicated in
Fig.~\ref{fig:effectiveArea}. For simplicity, we take $\varphi_\text{in} =
\varphi_\text{rfl} = \varphi_*$ and concentrate on the modulus of $\mathbf{k}$.
From (\ref{eq:trMr-start}) and (\ref{eq:func_f}), the Gaussian contribution of
a single reflection at a sphere with radius $R_j, j=1, 2$ can then be identified as
\begin{equation}
\exp\left(-\eta R_j\right) = \exp\left(-\frac{R_j}{4\kappa_*}
    (k_\text{in}-k_\text{rfl})^ 2\right)\,.
\label{eq:kGaussian}
\end{equation}
Here, $\eta$ is defined in analogy to (\ref{eq:eta}) with the two
wave-vector components replaced by $\kappa_\text{in}$ and $\kappa_\text{rfl}$.

Neglecting numerical factors of order one, the width around the
saddle-point manifold is thus $\delta k^{(j)} \sim \sqrt{\kappa_\ast/R_j}$. The
typical scale of $\kappa_*$ is set by the integral on the right-hand side of
(\ref{Mrp}), finally leading to the change of the projection of the wave
vector onto the $x$-$y$ plane
\begin{equation}
 \delta k^{(j)} = |\mathbf{k}_\text{rfl}-\mathbf{k}_\text{in}| \sim
    (LR_j)^{-1/2}\,.
 \label{eq:deltak}
\end{equation}
As expected, the scattering at the smaller sphere provides the larger
deviations from the saddle-point manifold. Thus, the effective area contributing
to the Casimir interaction is fixed by the smaller radius which we refer to
as $R_1$ in the following.

We now explore the implications of (\ref{eq:deltak}) in real space based on
specular reflection. The saddle-point manifold (\ref{eq:spmanifold}) corresponds
to reflections between the points on the two spheres corresponding to the
closest distance $L$ (cf.\ Fig.~\ref{fig:effectiveArea}). Deviations from the
saddle-point manifold as allowed by the Gaussian (\ref{eq:kGaussian}) implicate
the surroundings of these two points in the scattering process. We estimate the
dimension of the spherical cap on the surface of the smaller sphere 1 by
considering the scattering of propagating waves in the real frequency domain
with the help of Fig.~\ref{fig:effectiveArea}.

As above, we assume $\mathbf{k}_\text{in}$ and $\mathbf{k}_\text{rfl}$ to be
parallel and for simplicity omit the index $j=1$ when writing $\delta
k=|\mathbf{k}_\text{rfl}-\mathbf{k}_\text{in}|.$ As shown in Sec.~\ref{sec:wkb},
the WKB approximation for the direct reflection term amounts to a specular
reflection at a tangent plane $\pi(\vartheta)$ making an angle $\vartheta$ with
the $x$-$y$ plane. For parallel vectors and small values of $\delta k$, a
simple relation between $\vartheta$ and $\delta k$ can be derived by noting that
while the projection of the wave vector onto the plane $\pi(\vartheta)$ is
conserved during a scattering process, this is not the case for the projection
onto the $x$-$y$ plane for non-vanishing values of $\vartheta$. Assuming
$\vartheta\ll 1,$ we find
\begin{equation}
\delta k\approx 2\vartheta k_z\,.
\end{equation}
This relation together with the scaling $k_z\sim 1/L$ allow us to estimate the
width of the angular sector effectively contributing to the Casimir interaction
from the Gaussian width (\ref{eq:deltak}).

We find that the spherical cap around the point of closest distance corresponds
to the angular sector bounded by the angle $\vartheta \sim(L/R_1)^{1/2}\ll 1.$
As indicated by Fig.~\ref{fig:effectiveArea}, its transverse size is $d \approx
R_1\vartheta \sim (R_1L)^{1/2}\ll R_1<R_2.$ The same scaling was found by an
heuristic geometric argument \cite{Kim2010}. The area of the spherical surface
effectively contributing to the interaction is then $A \sim R_1L,$ which
coincides, except for a numerical factor of order one, with the ratio between
the Casimir force for two spheres within PFA and the Casimir pressure for
parallel planes, as long as the interaction obeys a power law \cite{footnote}.

Although the discussion above holds for arbitrary temperatures, we show in the
remainder of this section that the effective area for the thermal corrections
scales in a different way in the low-temperature regime.  The  difference arises
from the typical scale for $\kappa_*,$ which is no longer set by $1/L$, but
rather by $1/\lambda_T,$ where $\lambda_T = \hbar c/k_B T$ is the thermal
wavelength.  In order to illustrate this property, we consider the  thermal
correction of the Casimir force $\delta  F \equiv F(L, T)-F(L, 0)$ as an
example.  We start from Eqs.~(\ref{eq:PFA-result}) and (\ref{FPPLifshitz}) and
employ the Poisson summation formula \cite{Genet2000} to write
\begin{multline}\label{eq:correction}
\delta  F= 2\hbar \,R_{\rm eff}\,\sum_{m=1}^\infty
 \sum_{p}
\int_0^\infty d\xi
 \cos(m\lambda_T\xi/c) \int_{\xi/c}^\infty
\frac{d\kappa}{2\pi} \kappa \\
\times\log\left(1-r^{(1)}_p r^{(2)}_p e^{-2\kappa L} \right)\,.
\end{multline}
In the low-temperature limit, $L\ll \lambda_T,$ the exponential $\exp(-2\kappa
L)$ can be taken to be approximately constant and does not provide a cutoff for
the $\kappa$ integration in (\ref{eq:correction}). For instance, in the case of
plasma metals, the correction $\delta F$ can be written in terms of simple
integrals involving trigonometric functions of $m\lambda_T \kappa,$ which are
similar to the expressions derived for the Casimir pressure between parallel
planes in Ref.~\cite{Genet2000}. The derivation of the low-temperature limit of
(\ref{eq:correction}) for Drude metals is more involved \cite{Ingold2009}, but
$1/\lambda_T$ also provides the typical scale of  $\kappa$ in this case.

As a consequence, the effective area contributing to the thermal correction
$\delta F$  is found to be of the order of $A^{(T)}\sim R_1\lambda_T$
and thus much larger than the area relevant for $F(L,T)$, which is
dominated by the zero-temperature (vacuum) contribution in the low-temperature
limit.  This result is consistent with the numerical examples for a scalar field
presented in Ref.~\cite{Weber2010}.

The thermal correction to the Casimir force has been measured in the
plane-sphere geometry by employing a coated lens with $R_1=15.6 \, {\rm cm}$
\cite{Sushkov2011}. The results were analyzed with the help of the PFA, which
can be expected to provide an accurate description of the thermal correction if
the transverse size of the lens is much larger than $\sqrt{R_1\lambda_T}\sim
1\,\mathrm{mm}$. If our estimate valid for $L\ll \lambda_T$ applies to the
experiment where $L\lesssim 0.4\lambda_T$, we can conclude that the lens was
indeed of sufficient size.

In most Casimir force measurements, thermal corrections are typically very
small. Nevertheless, our estimation of an enlarged effective area is still
relevant for thermodynamic quantities vanishing in the zero-temperature limit,
in particular for the Casimir entropy.

\section{Conclusions}
\label{sec:conclusions}
For two spheres of arbitrary radii, we have derived the proximity force
approximation expression for the Casimir free energy as the leading asymptotic
result for distances between the spheres much smaller than their radii.  We have
made use of the WKB Mie scattering amplitudes where only the direct reflection
term in the Debye expansion contributes to leading order. The trace over a
number of round-trip matrices has been evaluated within the saddle-point
approximation. The saddle point corresponds to the conservation of the
wave-vector component perpendicular to the line connecting the centers of the
two spheres.  Therefore, the leading asymptotics results from specular
reflection in the vicinity of the points of closest distance between the
spheres. As an important consequence, we find that no polarization mixing
contributes to leading order.  The special case of the plane-sphere geometry is
recovered by taking the radius of one sphere to infinity.

Although our approach is based on the momentum representation, we are able to
estimate the effective area contributing to the Casimir interaction by making
use of the localization principle, which allows us to associate a specific
impact parameter to a given scattering angle in the WKB approximation.  Taken
together, the results presented here show that the PFA regime is governed by
local scattering from an area of the order of $R_1 L$ around the
points of closest approach, where $R_1$ is the radius of the smaller sphere. On
the other hand, for thermal corrections in the low-temperature regime, the area
becomes much larger and is of the order of $R_1 \lambda_T.$ As not all Casimir
experiments make use of whole spheres, these estimations provide a condition on
the minimum size of the spherical surface required for the PFA to hold for
the sphere-sphere or plane-sphere geometries. From a more theoretical
perspective, our results help to understand why local approaches such as the
derivative expansion are capable of providing both the leading and
next-to-leading order terms in several situations of interest.

\begin{acknowledgments}
We thank A. Lambrecht, S. Reynaud, and H. M. Nussenzveig for inspiring
discussions.  This work has been supported by CAPES and DAAD through the
PROBRAL collaboration program.  PAMN also thanks the Brazilian agencies
National Council for Scientific and Technological Development (CNPq), the
National Institute of Science and Technology Complex Fluids (INCT-FCx), the
Carlos Chagas Filho Foundation for Research Support of Rio de Janeiro
(FAPERJ) and the S\~ao Paulo Research Foundation (FAPESP).
\end{acknowledgments}

\appendix
\section{Derivation of the reflection operator matrix elements}
\label{sec:appendixA}

In the sphere-sphere geometry, the axis connecting the centers of the two
spheres is distinguished. We have defined it as $z$-axis and used the
polarization basis $\{\hat{\bm{\epsilon}}_\text{TE}, \hat{\bm{\epsilon}}_\text{TM}\}$
taken with respect to the incidence plane as specified in
(\ref{eq:polarizationTETM}). When a plane wave $\vert\mathbf{K}_i, p_i\rangle$
is scattered into a plane wave $\vert\mathbf{K}_j, p_j\rangle$,
the relations between the polarization basis vectors can be expressed as
\begin{equation}
\begin{aligned}
\hat{\bm{\epsilon}}_{\rm TE}({\bf K}_i) \cdot \hat{\bm{\epsilon}}_{\rm TE}({\bf K}_j)&=\phantom{-}\cos(\varphi)\\
\hat{\bm{\epsilon}}_{\rm TM}({\bf K}_i) \cdot \hat{\bm{\epsilon}}_{\rm TM}({\bf
K}_j)&=-\frac{c^2}{\xi^2}\left[k_ik_j-\phi_i\phi_j\kappa_i\kappa_j\cos(\varphi)\right]\\
\hat{\bm{\epsilon}}_{\rm TE}({\bf K}_i) \cdot \hat{\bm{\epsilon}}_{\rm TM}({\bf K}_j)&=-\frac{c\phi_j\kappa_j}{\xi}\sin(\varphi)\\
\hat{\bm{\epsilon}}_{\rm TM}({\bf K}_i) \cdot \hat{\bm{\epsilon}}_{\rm TE}({\bf K}_j)&=\phantom{-}\frac{c\phi_i\kappa_i}{\xi}\sin(\varphi)\,,
\label{eq:scalarproducts_inc}
\end{aligned}
\end{equation}
where $\varphi = \varphi_j-\varphi_i$.

Another distinguished polarization basis is defined by the scattering plane
spanned by the two wave vectors involved in the scattering process. In the
corresponding basis with polarization vectors perpendicular and parallel
to the scattering plane defined as
\begin{equation}
\begin{aligned}\label{eq:polarizationPerpPara}
\hat{\bm{\epsilon}}_\perp(\mathbf{K}_i) &= \frac{\hat{\mathbf{K}}_j \times \hat{\mathbf{K}}_i}
                                 {\vert\hat{\mathbf{K}}_j \times \hat{\mathbf{K}}_i\vert} \\
\hat{\bm{\epsilon}}_\parallel(\mathbf{K}_i) &= \hat{\bm{\epsilon}}_\perp \times \hat{\mathbf{K}}_i \,,
\end{aligned}
\end{equation}
respectively, the polarization is conserved during the scattering process.
More specifically, we have \cite{BH}
\begin{equation}
\begin{aligned}
\mathcal{R}\ket{\mathbf{K}_i,\perp}&=\frac{2\pi c}{\xi \kappa_j}S_1\ket{\mathbf{K}_j,\perp}\\
\mathcal{R}\ket{\mathbf{K}_i,\parallel}&=\frac{2\pi c}{\xi
\kappa_j}S_2\ket{\mathbf{K}_j,\parallel}
\end{aligned}
\end{equation}
with the scattering amplitudes $S_1$ and $S_2$ defined in (\ref{eq:SA}).
The basis vectors (\ref{eq:polarizationPerpPara}) for the incoming and
outgoing wave vectors are related by
\begin{equation}
\begin{aligned}
\hat{\bm{\epsilon}}_\perp(\mathbf{K}_i) \cdot \hat{\bm{\epsilon}}_\perp(\mathbf{K}_j) &= 1\\
\hat{\bm{\epsilon}}_\parallel(\mathbf{K}_i) \cdot \hat{\bm{\epsilon}}_\parallel(\mathbf{K}_j) &=\cos(\Theta)\\
\hat{\bm{\epsilon}}_\parallel(\mathbf{K}_i) \cdot \hat{\bm{\epsilon}}_\perp(\mathbf{K}_j)& = 0 \\
\hat{\bm{\epsilon}}_\perp(\mathbf{K}_i) \cdot \hat{\bm{\epsilon}}_\parallel(\mathbf{K}_j) &=0\,.
\end{aligned}
\end{equation}

The coefficients $A, B, C,$ and $D$ appearing in (\ref{eq:RS}) reflect the
change of polarization basis. They can be expressed as
\begin{equation}
\begin{aligned}
A &= \big(\hat{\bm{\epsilon}}_\text{TE}(\mathbf{K}_j)\cdot
          \hat{\bm{\epsilon}}_\perp(\mathbf{K}_j)\big)
     \big(\hat{\bm{\epsilon}}_\text{TE}(\mathbf{K}_i)\cdot
          \hat{\bm{\epsilon}}_\perp(\mathbf{K}_i)\big)\\
B &= \big(\hat{\bm{\epsilon}}_\text{TM}(\mathbf{K}_j)\cdot
          \hat{\bm{\epsilon}}_\perp(\mathbf{K}_j)\big)
     \big(\hat{\bm{\epsilon}}_\text{TM}(\mathbf{K}_i)\cdot
          \hat{\bm{\epsilon}}_\perp(\mathbf{K}_i)\big)\\
C &=-\big(\hat{\bm{\epsilon}}_\text{TM}(\mathbf{K}_j)\cdot
          \hat{\bm{\epsilon}}_\perp(\mathbf{K}_j)\big)
     \big(\hat{\bm{\epsilon}}_\text{TE}(\mathbf{K}_i)\cdot
          \hat{\bm{\epsilon}}_\perp(\mathbf{K}_i)\big)\\
D &= \big(\hat{\bm{\epsilon}}_\text{TE}(\mathbf{K}_j)\cdot
          \hat{\bm{\epsilon}}_\perp(\mathbf{K}_j)\big)
     \big(\hat{\bm{\epsilon}}_\text{TM}(\mathbf{K}_i)\cdot
          \hat{\bm{\epsilon}}_\perp(\mathbf{K}_i)\big)\,.
\end{aligned}
\end{equation}
Alternative expressions can be obtained by means of the relations
\begin{equation}
\begin{aligned}
\hat{\bm{\epsilon}}_\text{TE}\cdot\hat{\bm{\epsilon}}_\parallel &=
    -\hat{\bm{\epsilon}}_\text{TM}\cdot\hat{\bm{\epsilon}}_\perp\\
\hat{\bm{\epsilon}}_\text{TM}\cdot\hat{\bm{\epsilon}}_\parallel &=
    \hat{\bm{\epsilon}}_\text{TE}\cdot\hat{\bm{\epsilon}}_\perp\,.
\end{aligned}
\end{equation}

Expressing the scalar products (\ref{eq:scalarproducts_inc}) in terms of
the polarization basis $\{\hat{\bm{\epsilon}}_\perp, \hat{\bm{\epsilon}}_\parallel\}$,
we find the relations
\begin{equation}
\begin{aligned}
\hat{\bm{\epsilon}}_\text{TE}(\mathbf{K}_i)\cdot\hat{\bm{\epsilon}}_\text{TE}(\mathbf{K}_j)
&= A+B\cos(\Theta)\\
\hat{\bm{\epsilon}}_\text{TM}(\mathbf{K}_i)\cdot\hat{\bm{\epsilon}}_\text{TM}(\mathbf{K}_j)
&= A\cos(\Theta)+B\\
\hat{\bm{\epsilon}}_\text{TE}(\mathbf{K}_i)\cdot\hat{\bm{\epsilon}}_\text{TM}(\mathbf{K}_j)
&= -C-D\cos(\Theta)\\
\hat{\bm{\epsilon}}_\text{TM}(\mathbf{K}_i)\cdot\hat{\bm{\epsilon}}_\text{TE}(\mathbf{K}_j)
&= C\cos(\Theta)+D\,.
\end{aligned}
\end{equation}
Solving for the coefficients $A, B, C,$ and $D$ and making use of
(\ref{eq:scalarproducts_inc}), we finally obtain
\begin{widetext}
\begin{equation}
\begin{aligned}
A(\mathbf{K}_{i},\mathbf{K}_{j}) &=
\phantom{-}\frac{\xi^4\cos(\varphi)-c^4\big[k_ik_j\cos(\varphi)-\phi_i\phi_j\kappa_i\kappa_j\big]
\big[k_ik_j-\phi_i\phi_j\kappa_i\kappa_j\cos(\varphi)\big]}
{\xi^4-c^4\big[k_ik_j\cos(\varphi)-\phi_i\phi_j\kappa_i\kappa_j\big]^2}\\
B(\mathbf{K}_{i},\mathbf{K}_{j}) & = -\frac{\xi^2 c^2 k_i k_j\sin^2(\varphi)}
{\xi^4-c^4\big[k_ik_j\cos(\varphi)-\phi_i\phi_j\kappa_i\kappa_j\big]^2} \\
C(\mathbf{K}_{i},\mathbf{K}_{j}) & =\phantom{-} c^3\xi\sin(\varphi)\frac{k_ik_j\phi_i\kappa_i\cos(\varphi)-k_i^2\phi_j\kappa_j}
{\xi^4-c^4\big[k_ik_j\cos(\varphi)-\phi_i\phi_j\kappa_i\kappa_j\big]^2} \\
D(\mathbf{K}_i,\mathbf{K}_j) &= C(-\mathbf{K}_j,-\mathbf{K}_i)\,.
\end{aligned}
\label{eq:abcd}
\end{equation}
\end{widetext}
For $\xi=0$, they simplify to
\begin{equation}
A=-\phi_i\phi_j,\quad B=C=D=0\,,
\end{equation}
and for $\mathbf{k}_i=\mathbf{k}_j$ we find
\begin{equation}
A=1,\quad B=C=D=0\,.
\end{equation}
The matrix elements of the reflection operators \eqref{eq:RS} are not all mutually independent since
they fulfill reciprocity relations \cite{Carminati98, Messina2011,Messina2015}. In our notation these relations read
\begin{multline}
\kappa_i\bra{\mathbf{K}_i, p_i} \mathcal{R} \ket{\mathbf{K}_j,p_j} = \\
\kappa_j (-1)^{p_i+p_j}\bra{-\mathbf{K}_j, p_j} \mathcal{R} \ket{-\mathbf{K}_i,p_i}
\end{multline}
where $(-1)^{p_i+p_j}$ is $+1$ if the polarizations $p_i$ and $p_j$ are equal
and $-1$ otherwise. Indeed, since
\begin{equation}
\begin{aligned}
A(\mathbf{K}_i,\mathbf{K}_j) &= A(-\mathbf{K}_j,-\mathbf{K}_i)\\
B(\mathbf{K}_i,\mathbf{K}_j) &= B(-\mathbf{K}_j,-\mathbf{K}_i)\,,
\end{aligned}
\end{equation}
it is straightforward to verify that the coefficients (\ref{eq:abcd}) satisfy
the reciprocity relations.

\section{Low-frequency limit of the scattering amplitudes}
\label{sec:appendixB}

The WKB approximation for the scattering amplitudes (\ref{eq:SA-asymptotics})
at a sphere discussed in Sect.~\ref{sec:wkb} has been derived for large
size parameters $\xi R/c$.  As a consequence,
the zero-frequency contribution
 in the Matsubara sum
(\ref{eq:F}) is a priori  not covered by (\ref{eq:SA-asymptotics}).

By analyzing the low-frequency limit of the scattering amplitudes, we will
demonstrate that the scattering amplitudes (\ref{eq:SA-asymptotics}) and the
matrix elements (\ref{eq:RS}) obtained from them can be employed even in the
limit of zero frequency. At this point, it is worth noting that even though
the scattering amplitudes (\ref{eq:SA-asymptotics}) vanish in the limit $\xi\to0$,
this is not the case for the matrix elements (\ref{eq:RS}). Therefore, we need
to keep terms linear in $\xi$ in the low-frequency expression for the scattering
amplitudes.

In the following, we will distinguish three classes of materials: perfect reflectors,
real metals with a finite dc conductivity and dielectrics.
For perfect reflectors the permittivity is infinite for all frequencies, comprising
also the plasma model when the sphere radius is much larger than the plasma
wavelength. Real metals exhibit a finite permittivity except in the zero-frequency
limit where the finite dc conductivity gives rise to a divergence proportional
to $1/\xi$. Finally, for dielectrics, the permittivity remains finite for $\xi\to 0$.

We start from the expressions (\ref{eq:SA}) for the Mie scattering amplitudes and
first consider the material-independent functions $\pi_\ell$ and $\tau_\ell$. Noting
that according to (\ref{eq:costheta}), $\cos(\Theta)$ at low frequencies diverges
like $1/\xi^2$, we find the dominant low-frequency behavior
\begin{align}
\pi_\ell\big(\cos(\Theta)\big) &\simeq \frac{(2\ell)!}{2^\ell (\ell-1)! \ell!} \cos^{\ell-1}(\Theta)
	\sim\frac{1}{\xi^{2\ell-2}}\,, \\
\tau_\ell\big(\cos(\Theta)\big) &\simeq \frac{(2\ell)!}{2^\ell [(\ell-1)!]^2} \cos^{\ell}(\Theta)
	\sim\frac{1}{\xi^{2\ell}}\,.
\end{align}
As a consequence, among the four combinations of these two functions and the
two Mie coefficients in (\ref{eq:SA}), only those involving $\tau_\ell$ can
potentially lead to contributions linear in $\xi$. Terms involving $\pi_\ell$
yield an additional factor $\xi^2$ and can thus be disregarded. Furthermore,
only Mie coefficients going like $\xi^{2\ell+1}$ can then lead to a relevant
contribution to the scattering amplitudes.

Such a behavior is found for the Mie coefficient $a_\ell$ for which the leading term
at low frequencies can be expressed as
\begin{equation}
a_\ell \simeq (-1)^\ell \frac{(\ell+1)(\ell!)^2}{2\ell(2\ell+1)[(2\ell)!]^2}
	E^\mathrm{mat}_\ell\left(\frac{2\xi R}{c}\right)^{2\ell+1}\,.
\end{equation}
The material dependence is contained in the factor $E^\text{mat}_\ell$ (see
\cite{Canaguier-Durand11} for a detailed discussion). For dielectrics, one finds
\begin{equation}
E^\text{diel}_\ell = \frac{\varepsilon(0)-1}{\varepsilon(0)+\frac{\ell+1}{\ell}}\,,
\label{eq:ediel}
\end{equation}
while $E^\text{mat}_\ell=1$ for real metals and perfect reflectors. For the Mie coefficient
$b_\ell$, the required frequency dependence is only found for perfect reflectors where
\begin{equation}
b_\ell \sim - \frac{\ell}{\ell+1} a_\ell\,.
\end{equation}
In contrast, the low-frequency behavior of the Mie coefficient $b_\ell$ for real metals
is of order $\xi^{2\ell+2}$ and for dielectrics of order $\xi^{2\ell+3}$.

Let us now first consider the scattering amplitude $S_1(\Theta)$ which according
to the preceding analysis is only nonvanishing for perfect metals. This finding
is in agreement with the fact that for zero frequency the reflection
coefficient for TE modes vanishes for metals with finite dc conductivity and
dielectrics. Inserting the Mie coefficient $b_\ell$ and the function
$\tau_\ell(\cos\Theta)$ for perfect metals, we obtain
\begin{equation}
S_1(\Theta) \simeq -\frac{\xi R}{c}\sum_{\ell=1}^\infty\frac{\ell}{\ell+1}
\frac{\left(\frac{R}{c}\big(-2\xi^2\cos(\Theta)\big)^{1/2}\right)^{2\ell}}{(2\ell)!}
\end{equation}
We recall that in the matrix elements (\ref{eq:RS}) of the reflection operator
this function appears together with a prefactor $\xi^{-1}$. Furthermore, according
to (\ref{eq:costheta}), $\xi^2\cos(\Theta)$ does not vanish in the limit $\xi\to
0$. For large radius $R$, we then find
\begin{equation}
S_1(\Theta) \simeq -\frac{\xi R}{2c}e^{2(\xi R/c)\sin(\Theta/2)}
\label{eq:S1xi0}
\end{equation}
where we made use of the relation
\begin{equation}
 \left(-2\cos(\Theta)\right)^{1/2} = 2\sin(\Theta/2)
\end{equation}
valid for $\xi=0$.
For perfect reflectors, where $r_\text{TE}=-1$, (\ref{eq:S1xi0}) agrees with the
WKB result (\ref{eq:SA-asymptotics}).

The validity of the WKB approximation of $S_2(\Theta)$ can be proven along the same lines,
observing that for perfect reflectors and real metals, $r_\text{TM}=1$ at zero frequency.
For large $R$, the dominant contribution to the scattering amplitudes arises from large
angular momenta $\ell$. Therefore, $E^\text{diel}_\ell$ can be replaced by
$(\varepsilon(0)-1)/(\varepsilon(0)+1)$ which according to (\ref{eq:Fresnel-coeff}) agrees with
$r_\text{TM}$ in the low frequency limit. This completes the proof of the applicability
of (\ref{eq:SA-asymptotics}) even in the limit $\xi\to 0$.


\begin{thebibliography}{99}
\bibitem{footnote}
Since we have neglected numerical factors of order one, we are unable to
distinguish between the smallest radius $R_1$ and the effective radius $R_{\rm
eff}$ defined in (\ref{Reff}), given that their ratio is bounded by
$1<R_1/R_{\rm eff}\le 2.$
\end{thebibliography}
\end{document}